\documentclass[12pt,leqno]{article}
\usepackage{latexsym}
\newtheorem{theorem}{Theorem}
\newtheorem{corollary}{Corollary}
\newtheorem{lemma}{Lemma}

\newtheorem{definition}{Definition}

\newcommand{\blackslug}{\mbox{\hskip 1pt \vrule width 4pt height 8pt 
depth 1.5pt \hskip 1pt}}
\newcommand{\qed}{\quad\blackslug\lower 8.5pt\null\par\noindent}
\newenvironment{proof}{\par\noindent{\bf Proof:}}{\qed \par}

\newcommand{\cC}{\mbox{${\cal C}$}}
\newcommand{\cH}{\mbox{${\cal H}$}}
\newcommand{\cL}{\mbox{${\cal L}$}}

\newcommand{\Cn}{\mbox{${\cal C}n$}}

\newcommand{\eqdef}{\stackrel{\rm def}{=}}

\newcommand{\rev}[3]{\mbox{$#1$$#2$$#3$}}

\title{Algebras of Measurements: the logical structure of Quantum Mechanics
\thanks{This work was partially supported 
by the Jean and Helene Alfassa fund for 
research in Artificial Intelligence, by the Israel Science Foundation grant 
183/03 on ``Quantum and other cumulative logics'' and by EPSRC Visiting 
Fellowship GR/T 24562 on ``Quantum Logic''}
}

\author{Daniel Lehmann\\School of Engineering, 
\\Hebrew University, \\Jerusalem 91904, Israel  
\and Kurt Engesser \\Dept. of Computing, King's College, London
\and Dov M. Gabbay \\Dept. of Computing, King's College, London 
}
\date{July 2005, modified December 2005}

\begin{document}
\maketitle
\begin{abstract}
In Quantum Physics, a measurement is represented by a projection on some 
closed subspace of a Hilbert space. 
We study algebras of operators that abstract from the algebra of projections on
closed subspaces of a Hilbert space. 
The properties of such operators are justified 
on epistemological grounds. Commutation of measurements is a central topic
of interest. Classical logical systems may be viewed as measurement 
algebras in which all measurements commute.
\end{abstract}

\section{Introduction} \label{sec:intro}
We define a new class of abstract structures
for which we coin the term {\em algebras of measurements}, 
M-algebras for short. Those structures
are intended to capture the logic of physical measurements and in particular
of quantum measurements. From the physicist's point of view, it provides a 
framework, devoid of real physics and numbers, in which both classical and 
quantum mechanics can be described and the difference between them put in
evidence. Classical physics is, there, a special, very limited, almost trivial,
case of quantum physics. From the logician's point of view, it provides a 
generalized view of (non-monotonic) logic in which classical, i.e. monotonic,
logic is a special, very limited, almost trivial case.

This work takes its inspiration from the pioneering work of 
G. Birkhoff and J. von Neumann~\cite{BirkvonNeu:36}.
They proposed the view that {\em experimental propositions} 
are closed subspaces of a Hilbert
space and measurements are projections on such closed subspaces.
Strangely enough, they presented only a very preliminary analysis of the
properties of such projections and this topic seems to have been almost
ignored since. It had to wait for the thoroughly
new point of view proposed by Engesser and Gabbay in~\cite{EngGabbay:Quantum}.
In this paper, we take the following views:
\begin{itemize}
\item the Hilbertian formalization of Quantum Physics has been so extremely
successful for the reason that the algebra of projections in Hilbert spaces
possesses the properties that are epistemologically necessary to deal with
measurements that change the state of the system measured. We shall
therefore provide epistemological justifications to the properties possessed
by the Hilbertian formalism,
\item not all properties of the Hilbertian formalism are epistemologically
justified with the same force. Some (logical) aspects of Quantics may, 
probably, be studied, with advantage, in a weaker context,
\item the formalism of Measurement algebras suggests some, logically based, 
principles akin to superselection rules.
\end{itemize}

Let us, first, develop the analogy between measurements and propositions.
A physical measurement, e.g., measuring the temperature of a gas to be
$138{\,}^{\circ} K$, asserts that the proposition {\em the temperature of this 
gas is $138{\,}^{\circ} K$} holds true. A measurement, in a sense, asserts the
truth of a proposition. This is the fundamental analogy between physics and
logic: making a measurement is similar to asserting a certain kind of
proposition.
The example above has been taken from classical physics. Consider now 
measuring the spin of a particle along the z-axis to be $1 / 2$. 
This measurement is akin to asserting the truth of the proposition
{\em the spin along the z-axis is $1 / 2$}. But, here, the assertion of
the proposition, i.e., the measurement, changes the state of the system.
The assertion holds in the state resulting from the measurement, but did
not necessarily hold in the state of the system before the measurement
was performed. In fact it held in this previous state if and only if 
the measurement left the state unchanged. Inspired by the analogy 
between measurements and propositions we set ourselves to study the
logic of propositions that not only {\em hold}
at states, i.e., models, but also operate on them, 
transforming the state in which they are evaluated into another one. 
A proposition holds in some state if
and only if this state is a fixpoint for the proposition.

Section~\ref{sec:brief} will summarize in a most succinct and formal way
the definition of Algebras of Measurements (M-algebras), 
by presenting a list of
properties. 
It should be used as an overview and memento only.
The following sections will explain the properties, present motivation
and explanation, and then prove basic properties of M-algebras.
 
\section{M-algebras} \label{sec:brief}
The structures we are concerned with deal with a set $X$ and functions
from $X$ to $X$.
We shall denote the composition of functions by $\circ$ and composition 
has to be understood from left to right: for any \mbox{$x \in X$},
\mbox{$(\alpha \circ \beta)(x) =$} \mbox{$\beta(\alpha(x))$}.
If \mbox{$\alpha : X \longrightarrow X$}, we shall denote by 
\mbox{$FP(\alpha)$} the set of all fixpoints of $\alpha$:
\mbox{$FP(\alpha) \eqdef \{x \in X \mid \alpha(x) = x\}$}.

\begin{definition} \label{def:M-alg}
An M-algebra is a pair \mbox{$\langle X , M \rangle$} in which
$X$ is a non-empty set and $M$ is a family of functions from $X$ to $X$, 
that satisfies the six properties described below.
\end{definition}
\begin{itemize}
\item {\bf Illegitimate}  \mbox{$\exists \: 0 \in X$} such that 
\mbox{$\forall \alpha \in M$}, \mbox{$0 \in FP(\alpha)$}, i.e, 
\mbox{$\alpha(0) = 0$}.

\item {\bf Idempotence} \mbox{$\forall \alpha \in M$}, 
\mbox{$\alpha \circ \alpha = \alpha$}, i.e., for any \mbox{$x \in X$},
\mbox{$\alpha(\alpha(x)) = \alpha(x)$}.

The next property requires a preliminary definition.
\begin{definition} \label{def:preserves}
For any \mbox{$\alpha, \beta : X \longrightarrow X$}, we shall say that
$\alpha$ {\em preserves} $\beta$ if and only if $\alpha$ preserves
\mbox{$FP(\beta)$}, i.e., if \mbox{$\alpha(FP(\beta)) \subseteq FP(\beta)$},
i.e., \mbox{$\forall x \in X$}, 
\mbox{$\beta(x) = x \Rightarrow$} \mbox{$\beta(\alpha(x)) =$}
\mbox{$\alpha(x)$}.
\end{definition}

\item {\bf Composition} \mbox{$\forall \alpha, \beta \in M$}, if $\alpha$
preserves $\beta$, then \mbox{$\beta \circ \alpha \in M$}.

\item {\bf Interference} \mbox{$\forall x \in X$}, 
\mbox{$\forall \alpha, \beta \in M$}, if \mbox{$x \in FP(\alpha)$},
i.e., \mbox{$\alpha(x) = x$},
and \mbox{$(\beta \circ \alpha)(x) \in FP(\beta)$}, 
i.e., \mbox{$\beta(\alpha(\beta(x))) = \alpha(\beta(x))$},
then
\mbox{$\beta(x) \in FP(\alpha)$}, i.e., 
\mbox{$\alpha(\beta(x)) = \beta(x)$}.

\item {\bf Cumulativity} \mbox{$\forall x \in X$}, 
\mbox{$\forall \alpha, \beta \in M$},
if \mbox{$\alpha(x) \in FP(\beta)$}, 
i.e., \mbox{$\beta(\alpha(x)) = \alpha(x)$} and 
\mbox{$\beta(x) \in FP(\alpha)$}, 
i.e., \mbox{$\alpha(\beta(x)) = \beta(x)$}, then
\mbox{$\alpha(x) = \beta(x)$}.

The next property requires some notation.
For any \mbox{$\alpha : X \longrightarrow X$}, we shall denote by 
\mbox{$Z(\alpha)$} the set of zeros of $\alpha$:
\mbox{$Z(\alpha) \eqdef \{x \in X \mid \alpha(x) = 0\}$}.

\item {\bf Negation} \mbox{$\forall \alpha \in M$}, 
\mbox{$\exists (\neg \alpha) \in M$},
such that \mbox{$FP(\neg \alpha) =$} \mbox{$Z(\alpha)$},
 and \mbox{$Z(\neg \alpha) =$} \mbox{$FP(\alpha)$},
i.e., \mbox{$\forall x \in X$}, \mbox{$\alpha(x)=0$} iff 
\mbox{$(\neg \alpha)(x) = x$} and
\mbox{$\forall x \in X$}, \mbox{$\alpha(x)=x$} iff 
\mbox{$(\neg \alpha)(x) = 0$}.
\end{itemize}

Two additional properties will be considered in Section~\ref{sec:separable}.
\begin{definition} \label{def:separable}
An M-algebra is {\em separable} if it satisfies the following:
\item {\bf Separability}
For any \mbox{$x, y \in X - \{0\}$}, if \mbox{$x \neq y$}
then \mbox{$\exists \alpha \in M$} such that \mbox{$\alpha(x) = x$} and
\mbox{$\alpha(y) \neq y$}.
\end{definition}
\begin{definition} \label{def:strong_separable}
An M-algebra is {\em strongly-separable} if it satisfies the following:
\item {\bf Strong Separability}
For any \mbox{$x \in X - \{0\}$}, there exists a measurement 
\mbox{$e_{x} \in M$} such that
\mbox{$FP(e_{x}) = \{0 , x \}$}.
\end{definition}

\section{Motivation and Justification} \label{sec:motivation}
In this section, we shall leisurely explain each one of the properties
described in Section~\ref{sec:brief}.
Our explanation of each property will include three parts:
\begin{itemize}
\item an epistemological explanation whose purpose is to explain why the
property is natural or even required when one thinks of measurements,
\item an explanation of why the property holds in the algebra 
\mbox{$\langle H , L \rangle$} where $H$ is a Hilbert space and $L$ the set
of all projections onto closed subspaces of $H$,
\item an explanation of the logical meaning of the property, based on
the identification of measurements with propositions.
\end{itemize}

\subsection{States} \label{sec:states}
We shall reserve the term {\em state} for the elements of $X$.
In physical terms, the set $X$ is the set of all possible states of a system.
When we say {\em state} we mean a state as fully determined as is physically
possible: e.g., in classical mechanics, 
a set of $6 n$ values if we consider $n$ particles (three
values for position and three values for momentum), or what is generally 
termed, in Quantum Physics a {\em pure state}.

In the Hilbertian description of Quantum Physics, 
a (pure) state is a one-dimensional subspace, i.e., a ray,
in some Hilbert space. The illegitimate state, $0$ is the zero-dimensional
subspace.

A logician can give the term ``states'' two different acceptions.
It sometimes means a full definition of a possible state of the world,
i.e., what is true and what is false. 
This is a model, a maximal consistent theory.
Sometimes, it means a set of propositions known to be true. In this sense,
it is any consistent theory, not necessarily maximal.
We shall see that {\bf Illegitimate} forces us to consider the inconsistent
theory as a state.

\subsection{Measurements} \label{sec:measurements}
The elements of $M$ represent measurements on the physical system whose
possible states are those of $X$. 
In Classical Physics one may assume that a measurement leaves the 
measured system unchanged.
It is a hallmark of Quantum Physics that this assumption cannot be held
true anymore.
In Quantum Physics, measurements, in general, change the state of the system.
This is the phenomenon called {\em collapse of the wave function}.
Therefore we model measurements by transformations on the set of states.
Clearly not any transformation can be called a measurement. A measurement
changes the system in some minimal way. A transformation that brings about 
a wild change in the system cannot be considered to be a measurement.
Many of the properties presented above and discussed below explicit this
requirement.

A word of caution is necessary here before we proceed. When we speak
about measurement we do not mean some declaration of intentions such as 
{\em measuring the position of a particle}, we mean the action of measuring
some physical quantity {\em and finding a specific value}, such as {\em
finding the particle at the origin of the system of coordinates}.
Measuring $0.3{\,}^{\circ} K$ and measuring $1000{\,}^{\circ} K$ are not
two different possible results for the same measurement, they are two
different measurements.

In the Hilbertian description of Quantum Physics measurable quantities 
are represented by Hermitian operators. 
Measurements in our sense are represented by a pair
\mbox{$\langle A, \lambda \rangle$} where $A$ is a Hermitian operator and 
$\lambda$ an eigenvalue of $A$. 
The effect of measuring \mbox{$\langle A, \lambda \rangle$} in state $x$ is
to project $x$ onto the eigensubspace of $A$ for eigenvalue $\lambda$.
A measurement $\alpha$ is therefore a projection on a closed subspace of a 
Hilbert space. The set $FP(\alpha)$ is the closed subspace on which $\alpha$ 
projects.
Those projections onto eigensubspaces are the measurements we try to identify.
Our goal is to identify the algebraic properties of such projections
that make them suitable to represent physical measurements in Quantum
Physics.

From a classical logician's point of view, a measurement is a proposition.
A proposition $\alpha$ acts on a state, i.e., a theory $T$ by sending
it to the theory that results from adding $\alpha$ to $T$ and
then closing under logical consequence. One sees that, 
from this point of view, if $T$ is maximal then 
$\alpha(T)$ is either $T$ (iff $\alpha$ is in $T$) or
the inconsistent theory. We see here that a proposition (measurement) holds
in some model (state) if and only if the model is a fixpoint of the 
proposition. 

This is the interpretation that we shall take along with us: a measurement
$\alpha$ {\em holds} at some state $x$, or, equivalently $x$ satisfies
$\alpha$, if and only if 
\mbox{$x \in FP(\alpha)$}.

\subsection{Illegitimate} \label{sec:illegitimate}
{\bf Illegitimate} is mainly a technical requirement. The sequel will
show why it is handy.
The illegitimate state $0$ is a state that is physically impossible.
Physicists, in general, do not consider this state explicitly, we shall.
From the epistemological point of view, we just require that amongst all
the possible states of the system we include a state, denoted $0$ that
represents physical impossibility. There is not much sense in measuring 
anything in the illegitimate state, therefore, it is natural to assume 
that no measurement $\alpha$ operating on the
illegitimate state can change it into some legitimate state. 
This is the meaning of our requirement that $0$ be a fixpoint of any 
measurement. In other terms, the state $0$ satisfies every measurement,
every measurement holds at $0$.

In the Hilbertian description of Quantum Physics the zero vector plays
the role of our $0$. Indeed, since a projection is linear, it preserves
the zero vector.

From a logician's point of view {\bf Illegitimate} requires us to include
the inconsistent theory in $X$. Clearly, the result of adding any proposition
to the inconsistent theory leaves us with the inconsistent theory.

\subsection{Zeros} \label{sec:zeroes}
We have described in Section~\ref{sec:measurements} the interpretation we
give to the fact that a state $x$ is a fixpoint of a measurement $\alpha$.
We want to give a similarly central meaning to the fact that a state
$x$ is a zero of a measurement $\alpha$: \mbox{$x \in Z(\alpha)$}, i.e.,
\mbox{$\alpha(x) = 0$}. If measuring $\alpha$ sends $x$ to the illegitimate
state, measuring $\alpha$ is physically impossible at $x$. This should be
understood as meaning that the state $x$ has some definite value different
from the one specified by $\alpha$. 

If, at $x$, the spin is $1/2$ along the
z-axis, then measuring along the z-axis a spin of $- 1/2$ is physically 
impossible and therefore the measurement of $- 1/2$ sends the state $x$
to the illegitimate state $0$.
The status of the measurement that measures $- 1/2$ along the {\em $x$-axis}
is completely different: this measurement does not send $x$ to $0$, but to
some legitimate state in which the spin along the x-axis is $- 1/2$.

It is natural to say that a measurement $\alpha$ has a definite value
at $x$ iff $x$ is either a fixpoint or a zero of $\alpha$.
We shall define: \mbox{$Def(\alpha) \eqdef FP(\alpha) \cup Z(\alpha)$}.
If \mbox{$x \in Def(\alpha)$}, $\alpha$ has a definite value at $x$: either
it holds at $x$ or it is impossible at $x$.
If \mbox{$x \not \in Def(\alpha)$}, $\alpha(x)$ is some state different
from $x$ and different from $0$. 

In the Hilbertian presentation of Quantum Physics, the zeros of a measurement
$\alpha$ are the vectors orthogonal to the set of fixpoints of $\alpha$.

\subsection{Idempotence} \label{sec:idempotence}
{\bf Idempotence} is extremely meaningful.
It is an epistemologically fundamental property of measurements that they are
idempotent: if $\alpha$ is a measurement and $x$ a state,
then \mbox{$\alpha(\alpha(x)) = \alpha(x)$}, i.e., measuring the same value
twice in a row is exactly like measuring it once. 
Note that, by {\bf Illegitimate}, if \mbox{$x \in Def(\alpha)$}, then
\mbox{$\alpha(\alpha(x)) = \alpha(x)$}.
The import of {\bf Idempotence} concerns states that are not in $Def(\alpha)$.
 
It seems very difficult to imagine a scientific theory
in which measurements are not idempotent: it would be impossible to check
directly that a system is indeed in the state we expect it to be in without
changing it.
Idempotence is one of the conditions that ensure that measurements change
states only minimally.
This principle seems to be a fundamental
principle of all science, having to do with the reproducibility of
experiments. If there was a physical system and a measurement that, if
performed twice in a row gave different results, then such a measurement
would be, in principle, irreproducible.

In the Hilbertian description of Quantum Physics measurements are modeled 
by projections onto eigensubspaces. Any projection is idempotent.
But it is enlightening to reflect on the phenomenology of this
idempotence.
 For an electron whose
spin is positive along the $z$-axis (state $x_{0}$), 
measuring a negative spin along 
the $x$-axis is feasible, i.e., does not send the system into the illegitimate
state, but sends the system into a state ($x_{1}$)
different from the original one, $x_{0}$.
Nevertheless, a consequence of the collapse of the wave function is that,
after measuring a negative spin along the $x$-axis, the spin is indeed negative
along the $x$-axis and therefore a new measurement of a negative spin along 
the $x$-axis leaves the state $x_{1}$ of our electron unchanged, 
whereas measuring
a positive spin along the $x$-axis is now an unfeasible measurement and sends
$x_{1}$ to the illegitimate state. Note that such a measurement of a 
positive spin along the $x$-axis in the original state $x_{0}$ brings us to
a legitimate state $x_{3}$ different from $x_{0}$ and $x_{1}$.
The idempotence of measurements, probably epistemologically 
necessary, provides some explanation of why projections in Hilbert 
spaces are a suitable model.

From the logician's point of view, idempotence corresponds to the fact that
asserting the truth of a proposition is equivalent to asserting it twice.
For any reasonable consequence operation \cC, 
\mbox{$\cC(\cC(T, a), a) = \cC(T, a)$}.

\subsection{Preservation} \label{sec:preservation}
The definition of {\em preservation} encapsulates the way in which different
measurements can interfere. If $\alpha$ preserves $FP(\beta)$, 
the set of states in which $\beta$ holds, $\alpha$ never destroys 
the truth of proposition $\beta$: it never interferes badly with $\beta$.

\subsection{Composition} \label{sec:composition}
{\bf Composition} has physical significance.
It is a global principle: it assumes a global property and concludes a global
property.
Measurements are mappings of $X$ into itself, therefore we may consider
the composition of two measurements. According to the principle of minimal 
change, we do not expect the composition of two measurements to be a 
measurement: two small changes may make a big change. But, if those two
measurements do not interfere in any negative way with each other, 
we may consider their 
composition as small changes that do not add up to a big change.
{\bf Composition} requires that if, indeed, $\alpha$ preserves $\beta$,
then the composite operation that consists of measuring $\beta$ first,
and then $\alpha$ does not add up to a big change and should be a
bona fide measurement. Notice that we perform $\beta$ first, whose result is
(by {\bf Idempotence}) a state that satisfies $\beta$, 
then we perform $\alpha$,
which does not destroy the result obtained by the first measurement $\beta$.

In the Hilbertian presentation of Quantum Physics, consider $\alpha$, the
projection on some closed subspace $A$ and $\beta$, the projection on $B$.
The measurement $\alpha$ preserves $\beta$ iff the projection 
of the subspace $B$ onto $A$ is contained in the intersection $A \cap B$
of $A$ and $B$. 
In such a case
the composition $\beta \circ \alpha$ of the two projections, first on $B$
and then on $A$ is equivalent to the projection on the intersection $A \cap B$.
It is therefore a projection on some closed subspace.

For the classical logician, measurements always preserve each other.
If \mbox{$a \in T$}, then \mbox{$a \in \cC(T, b)$} for any proposition $b$.
This is a consequence of the monotonicity of \cC.
{\bf Composition} requires that the composition of any two measurements
be a measurement.
For the logician, \mbox{$\beta \circ \alpha$} is the measurement
\mbox{$\beta \wedge \alpha$}. 
{\bf Composition} amounts to the assumption that $M$ is closed under 
conjunction.

Technically, the role of {\bf Composition}is to ensure that two commuting 
measurements' composition is a measurement. Equivalently, we could have, 
instead of {\bf Composition}, required that for any pair 
\mbox{$\alpha, \beta \in M$} such that \mbox{$\alpha \circ \beta =$}
\mbox{$\beta \circ \alpha$}, their composition \mbox{$\alpha \circ \beta$} be
in $M$.

\subsection{Interference} \label{sec:Interference}
{\bf Interference} has a deep physical meaning. It is a local principle,
i.e., holds separately at each state $x$.
It may be seen as a local logical version of Heisenberg's uncertainty
principle. It considers a state $x$ that satisfies $\alpha$. 
Measuring $\beta$ at $x$ may leave $\alpha$ undisturbed 
(this is the conclusion), but, if $\beta$ disturbs $\alpha$, then no state
at which both $\alpha$ and $\beta$ hold can ever be attained by measuring
$\alpha$ and $\beta$ in succession. In other words, either such a state,
satisfying both $\alpha$ and $\beta$ is obtained immediately, or never.

We shall say that $\beta$ disturbs $\alpha$ at $x$ if \mbox{$x \in FP(\alpha)$}
but \mbox{$\beta(x) \not \in FP(\alpha)$}. Note that $\beta$ preserves 
$\alpha$ if and only if it disturbs $\alpha$ at no $x$.
{\bf Interference} says that if $\beta$ disturbs $\alpha$ at $x$ then
$\alpha$ disturbs $\beta$ at $\beta(x)$, and $\beta$ disturbs $\alpha$ at
\mbox{$(\beta \circ \alpha)(x)$}, and so on.
We chose to name this property {\em Interference} since it deals with the
local interference of two measurements: if they interfere once, they will
continue interfering ad infinitum.

In the Hilbertian presentation of Quantum Physics, the principle of 
{\bf Interference} is satisfied for the following reason.
Consider a vector \mbox{$x \in H$} and two closed subspaces of $H$: $A$ and
$B$. Assume $x$ is in $A$. Let $y$ be the projection of $x$ onto $B$ and
$z$ the projection of $y$ onto $A$. Assume that $z$ is in $B$.
Since both $x$ and $z$ are in $A$, the vector $z - x$ is in $A$.
Similarly, the vector $z - y$ is in $B$.
But $y$ is the projection of $x$ onto $B$ and therefore $y - x$ is orthogonal
to $B$ and in particular orthogonal to $z - y$.
We have \mbox{$(y - x) \cdot (z - y) = 0$}, and
\[
y \cdot z - y \cdot y - x \cdot z + x \cdot y = 0.
\]
Since $z$ is the projection of $y$ onto $A$, the vector $z - y$ is orthogonal 
to $A$ and we have \mbox{$(z - x) \cdot (z - y) = 0$}, and
\[
z \cdot z - z \cdot y - x \cdot z + x \cdot y = 0.
\]
By substracting the first equality from the second we get:
\[
- z \cdot y - y \cdot z + y \cdot y + z \cdot z =  
(y - z) \cdot (y - z) = 0.
\]
We conclude that \mbox{$y = z$}.

For the logician, it is always the case that \mbox{$\beta(x) \in FP(\alpha)$}
if \mbox{$x \in FP(\alpha)$}, as noticed in Section~\ref{sec:composition}.

\subsection{Cumulativity} \label{sec:cumulativity}
Cumulativity is motivated by Logic. It does not seem to have been reflected
upon by physicists. 
It parallels the cumulativity property that is central to nonmonotonic logic: 
see for example~\cite{KLMAI:89,Mak:Handbook,L:LogicsandSemantics}.
If the measurement of $\alpha$ at $x$ causes $\beta$ to hold (at $\alpha(x)$),
and the measurement of $\beta$ at $x$ causes $\alpha$ to hold (at $\beta(x)$)
then those two measurements have, locally (at $x$), the same effect.
Indeed, they cannot be directly distinguished by testing $\alpha$ and $\beta$.
{\bf Cumulativity} says that they cannot be distinguished even indirectly.

In the Hilbertian formalism, if the projection, $y$, of $x$ onto some closed
subspace $A$ is in $B$ (closed subspace) then $y$ is the projection of $x$
onto the intersection $A \cap B$. If the projection $z$ of $x$ onto $B$
is in $A$, $z$ is the projection of $x$ onto the intersection
$B \cap A$ and therefore $y = z$.
In fact, a stronger property than {\bf Cumulativity} holds in Hilbert spaces. 
The following property, similar to the Loop property of~\cite{KLMAI:89},
holds in Hilbert spaces:
{\bf L-Cumulativity} \mbox{$\forall x \in X$}, 
for any natural number $n$ and for any sequence
\mbox{$\alpha_{i}\in M$}, \mbox{$i = 0, \ldots , n$}
if, for any such $i$, 
\mbox{$\alpha_{i}(x) \in FP(\alpha_{i+1})$}, where 
$n + 1$ is understood as $0$, then, for any \mbox{$0 \leq i, j \leq n$},
\mbox{$\alpha_{i}(x) = \alpha_{j}(x)$}.

To see that this property holds in Hilbert spaces, consider the distance
$d_{i}$ between $x$ and the closed subspace $A_{i}$ on which $\alpha_{i}$
projects. The condition \mbox{$\alpha_{i}(x) \in FP(\alpha_{i+1})$}
implies that \mbox{$d_{i+1} \leq d_{i}$}. 
We have \mbox{$d_{0} \geq d_{1} \geq \ldots \geq d_{n} \geq d_{0}$} and
we conclude that all those distances are equal and therefore 
\mbox{$\alpha_{i}(x) \in FP(\alpha_{i+1})$} implies that 
\mbox{$\alpha_{i}(x) = \alpha_{i+1}(x)$}.
We do not know whether the stronger {\bf L-Cumulativity} is meaningful
for Quantum Physics, or simply an uninteresting consequence of the 
Hilbertian formalism.

For the logical point of view, one easily sees that any classical measurements
satisfy {\bf Cumulativity}, and even {\bf L-Cumulativity}.

\subsection{Negation} \label{sec:negation}
{\bf Negation} also originates in Logic. It corresponds to the assumption that
propositions are closed under negation. If $\alpha$ is a measurement, 
$\alpha$ tests whether a certain physical quantity has a specific value $v$.
If such a test can be performed, it seems that a similar test could be
performed to test the fact that the physical quantity of interest has
some other specific value or does not have value $v$.

In the Hilbertian formalism, to any closed subspace corresponds its 
orthogonal subspace, also closed.

For the logician, {\bf Negation} amounts to the closure of the set of
(classical) measurements, i.e., formulas, under negation.  

\subsection{Separability} \label{sec:separability}
We remind the reader that none of the separability assumptions is 
included in the defining properties of an M-algebra.
{\bf Separability} asserts that if any two non-zero states $x$ and $y$ 
are different, 
there is a measurement that holds at $x$ and not at $y$.
Indeed, if all measurements that hold at $x$ also hold at $y$ it would not
be possible to be sure that the system is in $x$ and not in $y$.
Compared to the previous requirements, 
{\bf Separability} is of quite a different kind. It is some akin
to a superselection principle, though presented in a dual way: a restriction
on the set of states not on the set of observables.

Note that this implies that, in any non-trivial M-algebra (an M-algebra is
trivial if \mbox{$X = \{0\}$} and \mbox{$M = \emptyset$}), every state
satisfies some measurement.

In the Hilbertian formalism, the projections on the one-dimensional 
subspaces defined by $x$ and $y$ respectively do the job.

For the logician, if $T_{1}$ and $T_{2}$ are two maximal consistent
sets that are different, there is a formula $\alpha$ in $T_{1} - T_{2}$.
But, one may easily find (non-maximal) different theories $T_{1}$ and $T_{2}$ 
such that \mbox{$T_{1} \subset T_{2}$}, contradicting {\bf Separability}.

{\bf Strong Separability} is a stronger requirement.
Indeed, in a strongly-separable M-algebra, for any non-zero, different, states 
$x , y$ the measurement $e_{x}$ holds at $x$ and not at $y$.

The epistemological motivation for such a strong requirement is the 
following. One must be able to prepare a system in each of its states, i.e., 
each of the non-zero elements $x$ of $X$.
One this has been done, one should be able to check that indeed the system 
is in the state it is claimed to be, i.e., there should be a measurement that
{\em measures} each non-zero state $x$: this measurement is $e_{x}$.

In the Hilbert space framework, every non-zero state is a one-dimensional 
subspace, therefore a closed subspace and a measurement. The same is true
in classical mechanics: every state of a system can be characterized by
one proposition stating all that is true about the system.

In the logical examples, we have seen above that considering all theories
as states defines an algebra that is not even separable. Considering
only maximal theories, on the contrary, provides for a strongly-separable
algebra, at least if there is only a finite set of atomic propositions, or
if we admit infinite conjunctions and disjunctions.
 
\section{Examples of M-algebras} \label{sec:examples}
In this section we shall formally define the two paradigmatical examples of
M-algebras that have been described in Section~\ref{sec:motivation}:
propositional calculus and Hilbert space.
\subsection{Logical Examples} \label{sec:Logicalexs}
\subsubsection{Propositional Calculus: a non-separable M-algebra 
and a separable one} 
\label{sec:PC}
We shall now formalize our treatment of Propositional Calculus as an M-algebra.
In doing so, we shall present Propositional Calculus in the way advocated
by Tarski and Gentzen.
Let \cL\ be any language closed under a unary connective $\neg$ and a binary
connective $\wedge$. Let \Cn\ be any consequence operation satisfying the
following conditions (the conditions are satisfied by Propositional Calculus).
\[
{\bf Inclusion} \ \forall A \subseteq \cL, \ A \subseteq \Cn(A),
\]
\[
{\bf Monotonicity} \ \forall A, B \subseteq \cL, A \subseteq B \ \Rightarrow
\ \Cn(A) \subseteq \Cn(B), 
\]
\[
{\bf Idempotence} \ \forall A \subseteq \cL, \Cn(A) = \Cn(\Cn(A)),
\]
\[
{\bf Negation} \ \forall A \subseteq \cL, a \in \cL, 
\Cn(A, \neg a) = \cL \ \Leftrightarrow \ a \in \cC(A), 
\]\[
{\bf Conjunction} \ \forall A \subseteq \cL, a, b \in \cL, 
\ \Cn(A, a, b) = \Cn(A, a \wedge b).
\]
Define a subset of \cL\ to be a 
{\em theory} iff it is closed under \Cn: \mbox{$T \subseteq \cL$}
is a theory iff \mbox{$\Cn(T) = T$}. Let $X$ be the set of all theories.
Let $M$ be the language \cL. The action of a formula \mbox{$\alpha \in \cL$}
on a theory $T$ is defined by: \mbox{$\alpha(T) = \Cn(T \cup \{\alpha\})$}.
In such a structure $\alpha$ holds at $T$ iff \mbox{$\alpha \in T$}.
Let us check that such a structure satisfies all the defining
properties of an M-algebra. We shall not mention the uses of Inclusion.
The illegitimate state is the theory \cL. {\bf Idempotence} 
follows from the property
of the same name. {\bf Composition} follows from {\bf Conjunction}: 
the composition \mbox{$a \circ b$} is the measurement \mbox{$a \wedge b$}.
Note that any pair of measurements commute.
{\bf Interference} is satisfied because \mbox{$a \in T$} implies 
\mbox{$a \in \Cn(T, b)$}.
{\bf Cumulativity} is satisfied because \mbox{$b \in \Cn(T, a)$} implies
\mbox{$\Cn(T, a) = \Cn(T, a, b)$} by {\bf Monotonicity} and {\bf Idempotence}.
{\bf Negation} holds by the property of the same name.
 
The M-algebra above does not satisfy 
{\bf Separability} since there are theories $T$ and $S$ such that 
\mbox{$T \subset S$} and every formula $\alpha$ satisfied by $T$ is also
satisfied by $S$. This M-algebra is {\em commutative}: 
any two measurements commute since:
\mbox{$\Cn(\cC(T, a), b) = \Cn(\cC(T, b), a)$}.

If we consider the subset \mbox{$Y \subset X$} consisting only of {\em maximal
consistent} theories and the inconsistent theory, we see that the pair
\mbox{$\langle Y , \cL \rangle$} is an M-algebra, because $Y$ is closed under
the measurements in \cL. In this M-algebra, all measurements do more than 
commute, they are {\em classical}, in the following sense.
\begin{definition} \label{def:classical}
A mapping \mbox{$\alpha : X \longrightarrow X$} 
is said to be {\em classical} iff for every
\mbox{$x \in X$}, either \mbox{$\alpha(x) = x$} or \mbox{$\alpha(x) = 0$}. 
\end{definition}
The M-algebra above is separable: if $T_{1}$ and $T_{2}$ are diferent 
maximal consistent theories there is a formula \mbox{$a \in T_{1} - T_{2}$}.
It is not strongly-separable, though, if there is no single 
formula equivalent to a maximal theory.

\subsubsection{Nonmonotonic inference operations} \label{sec:nonmon}
In Section~\ref{sec:PC} we assumed that the inference operation \Cn\ 
was monotonic. It seems attractive to consider the more general case
of nonmonotonic inference operations studied, for example 
in~\cite{L:LogicsandSemantics}.
More precisely what about replacing {\bf Monotonicity} by the weaker
\[
{\bf Cumulativity} \ \forall A, B \subseteq \cL,
\ A \subseteq B \subseteq \cC(A) \ \Rightarrow \ \cC(B) = \cC(A).
\]
Notice that, in such a case, we prefer to denote our inference operation 
by \cC\ and not by \Cn.
The reader may verify that all requirements for an M-algebra still hold true,
{\em except for {\bf Composition}}. In such a structure all measurements
still commute and we therefore need that every composition $a \circ b$ of
measurements (formula) be a measurement (formula). 
But the reader may check that $a \wedge b$ does not have the required
properties: \mbox{$\cC(T, a \wedge b) = \cC(T, a, b)$} but, since \cC\ is
not required to be monotonic, there may well be some formula 
\mbox{$c \in \cC(T, a)$} that is not in $\cC(T, a, b)$.
In such a case \mbox{$\cC(T, a \wedge b) \neq \cC(\cC(T,a), b)$},
as would be required.
One may, then, think of extending the language \cL\ to include formulas
of the form \mbox{$a \circ b$} acting as compositions. But the 
{\bf Negation} condition of the definition of an M-algebra requires every
formula (measurement) to have a negation and there is no obvious definition
for the negation of a composition. 
The {\bf Monotonicity} property seems therefore essential.

\subsubsection{Revisions} \label{sec:revisions}
Another natural idea is to consider revisions a la AGM~\cite{AGM:85}.
The action of a formula $a$ on a theory $T$ would be defined as the theory
$T$ revised by $a$: $\rev{T}{*}{a}$.
The structure obtained does not satisfy the M-algebra assumptions.
The most blatant violation concerns {\bf Negation}. In revision theory
negation does not behave at all as expected in an M-algebra. 

\subsection{Orthomodular and Hilbert spaces} \label{sec:orthoandHilbert}

J. von Neumann's~\cite{vonNeumann:Quanten} firmly set Quantum Mechanics in 
the framework of Hilbert spaces. We assume the definition of a Hilbert space
is known to the reader. Hilbert spaces are orthomodular spaces. We shall not
burden the reader with the definition of such spaces here: the reader may
replace, in the sequel, the word {\em orthomodular} by {\em Hilbert} and lose
little of the strength of the results. A fundamental (but not used in this 
paper) result of Sol\`{e}r~\cite{Soler:95} characterizes infinite-dimensional
Hilbert spaces in orthomodular spaces.

\subsubsection{Orthomodular spaces} \label{sec:ortho}
Given any orthomodular space \cH, denote by $M$ the set of all closed subspaces
of \cH. Then the pair \mbox{$\langle \cH , M \rangle$} is an M-algebra, if
any \mbox{$\alpha \in M$} acts on \cH\ in the following way:
$\alpha(x)$ is the unique vector such that \mbox{$x = \alpha(x) + y$} for
some vector \mbox{$y \in \alpha^{\bot}$}. In light of 
Section~\ref{sec:motivation} the reader will have no trouble proving that
any such structure is an M-algebra. It is not separable, though: any two
colinear vectors satisfy exactly the same measurements. The next section
will present a related separable M-algebra.

\subsubsection{Rays} \label{sec:rays}
Given any orthomodular space \cH, let $X$ be the set of one-dimensional or 
zero-dimensional subspaces of \cH. Let $M$ be the set of closed subspaces
of \cH. The projection on a closed subspace is linear and therefore sends
a one-dimensional subspace to a one-dimensional or a zero-dimensional 
subspace and sends the zero-dimensional subspace to itself. The pair
\mbox{$\langle X , M \rangle$} is easily seen to be an M-algebra.
This M-algebra is separable: notice that \mbox{$X \subset M$} and that 
\mbox{$x \in X$} is the only state satisfying the measurement $x$.

\section{Properties of M-algebras} \label{sec:properties}
We assume that \mbox{$\langle X , M \rangle$} is an arbitrary M-algebra.
First, we shall show that any M-algebra includes two trivial measurements:
$\top$, analogous to the truth-value {\em true}, 
that leaves every state unchanged and measures a property satisfied
by every state and $\bot$, analogous to {\em false}, that sends every state
to the illegitimate state, and is nowhere satisfied. 
\begin{lemma} \label{le:zeroid} [Negation, Composition, Idempotence]
There are measurements \mbox{$\top, \bot \in M$} such that for every
\mbox{$x \in X$}, \mbox{$\top(x) = x$} and \mbox{$\bot(x) = 0$}.
\end{lemma}
\begin{proof}
The set $M$ of measurements is not empty: assume \mbox{$\alpha \in M$}.
Clearly, by {\bf Negation}, the measurement $\neg \alpha$ preserves $\alpha$.
It follows, by {\bf Composition}, that \mbox{$\alpha \circ (\neg \alpha)$}
is a measurement. Let \mbox{$\bot = \alpha \circ (\neg\alpha)$}.
By {\bf Idempotence} and {\bf Negation},
for every \mbox{$x \in X$}, \mbox{$\bot(x) = 0$}. 
We now let \mbox{$\top = \neg \bot$}.
\end{proof}

Then, we want to show that measurements are uniquely specified by their
fixpoints.
\begin{lemma} \label{le:spec}[Idempotence, Cumulativity]
For any \mbox{$\alpha, \beta \in M$}, if \mbox{$FP(\alpha) = FP(\beta)$},
then \mbox{$\alpha = \beta$}.
\end{lemma}
\begin{proof}
Assume \mbox{$FP(\alpha) = FP(\beta)$}.
Let \mbox{$x \in X$}.
By {\bf Idempotence} 
\mbox{$\alpha(x) \in$} \mbox{$FP(\alpha)$} and therefore, by assumption
\mbox{$\alpha(x) \in$} \mbox{$FP(\beta)$}. 
Similarly \mbox{$\beta(x) \in$} \mbox{$FP(\alpha)$}.
By {\bf Cumulativity}, then, \mbox{$\alpha =$} \mbox{$\beta$}.
\end{proof}

\begin{corollary} \label{le:double_neg}[Idempotence, Cumulativity, Negation]
For any \mbox{$\alpha \in M$}, \mbox{$\neg \neg \alpha =$} \mbox{$\alpha$}.
\end{corollary}
\begin{proof}
Both $\alpha$ and $\neg \neg \alpha$ are measurements and 
\mbox{$FP(\neg \neg \alpha) =$} \mbox{$FP(\alpha)$}.
\end{proof}

We shall now prove a very important property. 
Suppose $x$ is a state in which some measurement
(i.e., proposition) holds: for example, at $x$ the spin along the $x$-axis 
is $1/2$.
Performing a measurement $\alpha$ on $x$ may lead to a different state
\mbox{$y = \alpha(x)$}. At $y$ the spin along the $x$-axis may still be $1/2$, 
or it may be the case that the measurement $\alpha$ has interfered with the
value of the spin. But, under no circumstance, can it be the case 
that the spin along the $x$-axis has a definite value different from $1/2$, 
such as $-1/2$. If the value of the spin along the $x$-axis at $y$ 
is not $1/2$, the spin must be indefinite.
This expresses the fact that a measurement
$\alpha$, acting on a state in which $\beta$ holds, can either preserve $\beta$
(when \mbox{$\alpha(x) \in FP(\beta)$}) or can disturb $\beta$ 
(when \mbox{$\alpha(x) \not \in Def(\beta)$}) 
but cannot make $\beta$ impossible
at $x$, i.e., \mbox{$\alpha(x) \in Z(\beta)$}.
This is a very natural requirement stemming from the {\em minimal change}
principle. A move from a definite value to a different definite value is
too drastic to be accepted as measurement.

In the Hilbertian presentation of Quantum Physics, 
measurements are projections. The projection of a non-null vector $x$ 
onto a closed subspace $A$ is never orthogonal to $x$, unless $x$ is orthogonal
to $A$. Therefore if $x$ is in some subspace $B$, but its projection on $A$
is orthogonal to $B$, then this projection is the null vector.
  
\begin{lemma}\label{le:definiteness} [Illegitimate, Interference] 
For any \mbox{$x \in X$}, 
\mbox{$\alpha, \beta \in M$}, if \mbox{$x \in FP(\beta)$},
i.e., \mbox{$\beta(x) = x$},
and \mbox{$\alpha(x) \in Z(\beta)$}, i.e., 
\mbox{$\beta(\alpha(x)) = 0$},
then \mbox{$x \in Z(\alpha)$}, i.e., \mbox{$\alpha(x) = 0$}.
\end{lemma}
\begin{proof}
Assume \mbox{$x \in FP(\beta)$} and \mbox{$\beta(\alpha(x)) = 0$}.
Then \mbox{$(\alpha \circ \beta)(x) = 0 \in FP(\alpha)$}.
By {\bf Interference}, then, \mbox{$\alpha(x) \in FP(\beta)$} and
\mbox{$\beta(\alpha(x)) = \alpha(x)$}, i.e.,
\mbox{$0 = \alpha(x)$}.
\end{proof}

We shall now sort out the relation between fixpoints and zeros.
The next result is a dual of Lemma~\ref{le:definiteness}.
\begin{lemma} \label{le:Def_dual}[Illegitimate, Interference, Negation]
\mbox{$\forall x \in X$}, 
\mbox{$\forall \alpha, \beta \in M$}, if \mbox{$x \in$} \mbox{$Z(\beta)$}
and \mbox{$\alpha(x) \in FP(\beta)$}, then \mbox{$x \in$} \mbox{$Z(\alpha)$}.
In other terms, if \mbox{$\beta(x) = 0$} and 
\mbox{$\beta(\alpha(x)) =$} \mbox{$\alpha(x)$}, then \mbox{$\alpha(x) = 0$}.
\end{lemma}
\begin{proof}
Consider the measurement $\neg \beta$ guaranteed by {\bf Negation}.
If we have \mbox{$x \in FP(\neg \beta)$}
and \mbox{$\alpha(x) \in Z(\neg \beta)$}, then,
by Lemma~\ref{le:definiteness} we have \mbox{$x \in Z(\alpha)$}.
\end{proof}

\begin{lemma} \label{le:FPZ} [Illegitimate, Idempotence, Interference, Negation]
For any \mbox{$\alpha, \beta \in M$},
\mbox{$FP(\alpha) \subseteq FP(\beta)$} iff 
\mbox{$Z(\beta) \subseteq Z(\alpha)$}.
\end{lemma}
\begin{proof}
Suppose \mbox{$FP(\alpha) \subseteq FP(\beta)$} and \mbox{$x \in Z(\beta)$}.
Since, by {\bf Idempotence}, \mbox{$\alpha(x) \in FP(\alpha)$},
we have, by assumption, \mbox{$\alpha(x) \in FP(\beta)$}.
By Lemma~\ref{le:Def_dual}, then \mbox{$x \in Z(\alpha)$}.

Suppose now that \mbox{$Z(\beta) \subseteq Z(\alpha)$}.
We have \mbox{$FP(\neg \beta) \subseteq FP(\neg \alpha)$}
and by what we just proved: \mbox{$Z(\neg \alpha) \subseteq Z(\neg \beta)$}.
We conclude that \mbox{$FP(\alpha) \subseteq FP(\beta)$}.
\end{proof}

We shall now consider the composition of measurements.
First we show the symmetry of the preservation relation.
\begin{lemma} \label{le:symm} [Idempotence, Interference]
For any \mbox{$\alpha, \beta \in M$}, $\alpha$ preserves $\beta$ iff
$\beta$ preserves $\alpha$.
\end{lemma}
\begin{proof}
Assume $\alpha$ preserves $\beta$, and \mbox{$x \in FP(\alpha)$}.
By {\bf Idempotence}, \mbox{$\beta(x) \in$} \mbox{$FP(\beta)$}.
Since $\alpha$ preserves $\beta$, 
\mbox{$\alpha(\beta(x)) \in$} \mbox{$FP(\beta)$}.
The assumptions of {\bf Interference} are satisfied and we conclude
that \mbox{$\beta(x) \in$} \mbox{$FP(\alpha)$}. We have shown that $\beta$ 
preserves $\alpha$.
\end{proof}

\begin{lemma} \label{le:comp_incl}[Illegitimate, Idempotence, Interference, Negation]
For any \mbox{$\alpha, \beta \in M$}, if \mbox{$\alpha \circ \beta \in M$},
then \mbox{$FP(\alpha \circ \beta) =$} \mbox{$FP(\alpha) \cap FP(\beta)$}.
\end{lemma}
\begin{proof}
Since \mbox{$Z(\alpha) \subseteq$} \mbox{$Z(\alpha \circ \beta)$}, 
Lemma~\ref{le:FPZ}
implies that \mbox{$FP(\alpha \circ \beta) \subseteq$} \mbox{$FP(\alpha)$}.
By {\bf Idempotence} of $\beta$, 
\mbox{$FP(\alpha \circ \beta) \subseteq$} \mbox{$FP(\beta)$}.
We see that 
\mbox{$FP(\alpha \circ \beta) \subseteq$} \mbox{$FP(\alpha) \cap FP(\beta)$}.
But the inclusion in the other direction is obvious.
\end{proof}

We shall now show that the converse of {\bf Composition} holds.
\begin{lemma} \label{le:preserves}[Illegitimate, Idempotence, Interference, Negation]
For any \mbox{$\alpha, \beta \in M$}, if \mbox{$\alpha \circ \beta \in M$},
then $\beta$ preserves $\alpha$.
\end{lemma}
\begin{proof}
By Lemma~\ref{le:comp_incl}, 
\mbox{$FP(\alpha \circ \beta) \subseteq FP(\alpha)$}.
For any $x$, \mbox{$(\alpha \circ \beta)(x)$} is therefore 
a fixpoint of $\alpha$.
Assume \mbox{$x \in FP(\alpha)$}.
Then, \mbox{$(\alpha \circ \beta)(x) = \beta(x)$} is a fixpoint of $\alpha$.
\end{proof}

\begin{lemma} \label{le:iffpres}[Illegitimate, Idempotence, Interference, Composition, Negation]
For any \mbox{$\alpha, \beta \in M$}, \mbox{$\alpha \circ \beta \in M$},
iff $\beta$ preserves $\alpha$.
\end{lemma}
\begin{proof}
The {\em only if} part is Lemma~\ref{le:preserves}.
The {\em if} part is {\bf Composition}.
\end{proof}

\begin{lemma} \label{le:iffcomp} [Illegitimate, Idempotence, Interference, Composition, Negation]
For any \mbox{$\alpha, \beta \in M$}, \mbox{$\alpha \circ \beta \in M$}
iff \mbox{$\beta \circ \alpha \in M$}.
\end{lemma}
\begin{proof}
By Lemmas~\ref{le:iffpres} and~\ref{le:symm}.
\end{proof}

\begin{lemma} \label{le:commutation} [Illegitimate, Idempotence, Interference, Composition, Cumulativity, Negation]
For any \mbox{$\alpha, \beta \in M$}, \mbox{$\alpha \circ \beta \in M$}
iff $\alpha$ and $\beta$ commute, i.e., 
\mbox{$\alpha \circ \beta = \beta \circ \alpha$}.
\end{lemma}
\begin{proof}
Assume, first, that \mbox{$\alpha \circ \beta \in M$}.
By Lemma~\ref{le:iffcomp}, \mbox{$\beta \circ \alpha \in M$}.
By Lemma~\ref{le:comp_incl},
\mbox{$FP(\alpha \circ \beta) = FP(\beta \circ \alpha)$},
which implies the claim by Lemma~\ref{le:spec}.

Assume, now that $\alpha$ and $\beta$ commute.
We claim that $\alpha$ preserves $\beta$: indeed,
if \mbox{$\beta(x) = x$}, then 
\mbox{$\beta(\alpha(x)) = \alpha(\beta(x)) = \alpha(x)$} and therefore,
by {\bf Composition}, \mbox{$\beta \circ \alpha$} is a measurement.
\end{proof}

\begin{lemma} \label{le:incl}[Illegitimate, Idempotence, Interference, Composition, Cumulativity, Negation]
For any \mbox{$\alpha, \beta \in M$}, 
if \mbox{$FP(\alpha) \subseteq FP(\beta)$}, then
\mbox{$\alpha \circ \beta = \beta \circ \alpha = \alpha$}.
\end{lemma}
\begin{proof}
If \mbox{$FP(\alpha) \subseteq FP(\beta)$}, then, clearly
\mbox{$\alpha \circ \beta = \alpha$} by Idempotence of $\alpha$.
Therefore \mbox{$\alpha \circ \beta \in M$} and, 
by Lemma~\ref{le:commutation}, $\alpha$ and $\beta$ commute.
\end{proof}

\section{Connectives in M-algebras} \label{sec:connectives}
\subsection{Connectives for arbitrary measurements} \label{sec:conn_arbitrary}
The reader has noticed that {\em negation} plays a central role in our 
presentation of M-algebras, through the {\bf Negation} requirement and that
this requirement is central in the derivation of many of the lemmas of
Section~\ref{sec:properties}. Indeed, {\bf Negation} expresses the 
orthogonality structure so fundamental in orthomodular and Hilbert spaces.
The requirement of {\bf Negation} corresponds, for the logician, to the
existence of a connective whose properties are those of a classical negation.
Indeed, for example, as shown by Corollary~\ref{le:double_neg}, 
double negations may be ignored, as is the case in classical logic.
In~\cite{BirkvonNeu:36}, the logical language presented includes negation,
interpreted as orthogonal complement, and this is consistent with our 
interpretation. But~\cite{BirkvonNeu:36} also defines other connectives:
conjunction, disjunction and many later works on Quantum Logic define also
implication (sometimes a number of implications).
Our treatment does not require such connectives, or more precisely,
our treatment does not require that such connectives be defined between any
pair of measurements. 

Consider conjunction. One may consider only M-algebras in which, for any
\mbox{$\alpha, \beta \in M$} there is a measurement 
\mbox{$\alpha \wedge \beta \in M$} such that 
\mbox{$FP(\alpha \wedge \beta) =$}
\mbox{$FP(\alpha) \cap FP(\beta)$}.
There are many such M-algebras, since any M-algebra defined by an
orthomodular space
and the family of all its (projections on) closed subspaces has this property
since the intersection of any two closed subspaces is a closed subspace.
But our requirements do not imply the existence of such a measurement
\mbox{$\alpha, \beta \in M$} for every $\alpha$ and $\beta$.

For disjunction, one may consider requiring that for any 
\mbox{$\alpha, \beta \in M$} there be a measurement 
\mbox{$\alpha \vee \beta \in M$} such that
\mbox{$Z(\alpha \vee \beta) = Z(\alpha) \cup Z(\beta)$},
and the M-algebras defined by Hilbert spaces satisfy this requirement.
Not all M-algebras satisfy this requirement.

For implication, in general M-algebras, assuming conjunction and disjunction,
one could require that for any 
\mbox{$\alpha, \beta \in M$} there be a measurement 
\mbox{$\alpha \rightarrow \beta \in M$} such that
\mbox{$FP(\alpha \rightarrow \beta) =$}
\mbox{$FP(\neg \alpha \vee (\alpha \wedge \beta))$},
and the M-algebras defined by Hilbert spaces satisfy this requirement.
Indeed, works in Quantum Logic sometimes consider more than one implication, 
see~\cite{DallaChiara:01}.

The thesis of this paper is that connectives should not be defined
for arbitrary measurements, but only for commuting measurements.
One of the novel features of M-algebras is that conjunction, 
disjunction and implication are defined only for commuting measurements.
The next section will show that this restriction leads to a classical 
propositional logic. If one restricts oneself to commuting measurements,
then, contrary to the unrestricted connectives of Birkhoff and von 
Neumann~\cite{BirkvonNeu:36}, conjunction and disjunction distribute,
and, in fact, the logic obtained is classical.


\subsection{Connectives for commuting measurements} 
\label{sec:conn_commuting}
Let us take a second look at propositional connectives in M-algebras,
with particular attention to their commutation properties.
We shall assume that \mbox{$\langle X , M \rangle$} is an M-algebra.

\subsubsection{Negation} \label{sec:conn_negation}
{\bf Negation} asserts the existence of a negation for every measurement.
Let us study the commutation properties of $\neg \alpha$.
\begin{lemma} \label{le:comm_neg}
\mbox{$\forall \alpha, \beta \in M$}, if $\alpha$ commutes with $\beta$,
then $\neg \alpha$ commutes with $\beta$.
\end{lemma}
\begin{proof}
Assume $\alpha$ commutes with $\beta$. 
We shall see that $\beta$ preserves $\neg \alpha$.
Let \mbox{$x \in FP(\neg \alpha)$}. We have \mbox{$x \in Z(\alpha)$}.
But \mbox{$(\alpha \circ \beta)(x) = (\beta \circ \alpha)(x)$}.
Therefore \mbox{$0 = \alpha(\beta(x))$}, \mbox{$\beta(x) \in Z(\alpha)$}
and \mbox{$\beta(x) \in FP(\neg \alpha)$}. We have shown that
$\beta$ preserves $\neg \alpha$.
By {\bf Composition}, \mbox{$(\neg \alpha) \circ \beta \in M$} and,
by Lemma~\ref{le:commutation}, $\neg \alpha$ commutes with $\beta$.
\end{proof}

\begin{corollary} \label{le:co_neg}
\mbox{$\forall \alpha, \beta \in M$},
$\alpha$ and $\beta$ commute iff $\neg \alpha$ and $\beta$ commute iff
$\alpha$ and $\neg \beta$ commute iff $\neg \alpha$ and $\neg \beta$ commute.
\end{corollary}
\begin{proof}
By Lemma~\ref{le:comm_neg} and Corollary~\ref{le:double_neg}.
\end{proof}
\subsubsection{Conjunction} \label{sec:conjunction}
We shall now define a conjunction between {\em commuting} measurements.
\begin{definition} \label{def:conj}
For any {\em commuting}
measurements \mbox{$\alpha, \beta \in M$}, the conjunction 
\mbox{$\alpha \wedge \beta$} is defined by:
\mbox{$\alpha \wedge \beta =$} \mbox{$\alpha \circ \beta =$} 
\mbox{$\beta \circ \alpha$}.
\end{definition}
By Lemma~\ref{le:commutation}, the conjunction, as defined, is indeed
a measurement.
\begin{lemma} \label{le:unique_conj}
For any commuting \mbox{$\alpha, \beta \in M$}, the conjunction 
\mbox{$\alpha \wedge \beta$} is the unique measurement $\gamma$
such that \mbox{$FP(\gamma) =$}
\mbox{$FP(\alpha) \cap FP(\beta)$}.
\end{lemma}
\begin{proof}
By Lemmas~\ref{le:spec} and~\ref{le:comp_incl}.
\end{proof}
One immediately sees that conjunction among commuting measurements
is associative, commutative and that 
\mbox{$\alpha \wedge \alpha = \alpha$} for any \mbox{$\alpha \in M$}. 

Let us now study the commutation properties of conjunction.
\begin{lemma} \label{le:comm_conj}
\mbox{$\forall \alpha, \beta, \gamma \in M$}, that commute in pairs,
\mbox{$\alpha \wedge \beta$} commutes with $\gamma$.
\end{lemma}
\begin{proof}
\[
(\alpha \wedge \beta) \circ \gamma =
(\alpha \circ \beta) \circ \gamma =
\alpha \circ (\beta \circ \gamma) =
\alpha \circ (\gamma \circ \beta) =
\]
\[
(\alpha \circ \gamma) \circ \beta =
(\gamma \circ \alpha) \circ \beta =
\gamma \circ (\alpha \circ \beta) =
\gamma \circ (\alpha \wedge \beta)
\]
\end{proof}

\subsubsection{Disjunction} \label{sec:disj}
One may now define a disjunction between two commuting measurements
in the usual, classical, way.
\begin{definition} \label{def:disj}
For any {\em commuting}
measurements \mbox{$\alpha, \beta \in M$}, the disjunction 
\mbox{$\alpha \vee \beta$} is defined by:
\mbox{$\alpha \vee \beta =$} \mbox{$\neg ( \neg \alpha \wedge \neg \beta)$}. 
\end{definition}
By Corollary~\ref{le:co_neg}, the measurements
$\neg \alpha$ and $\neg \beta$ commute, therefore their conjunction is
well-defined and the definition of disjunction is well-formed.

The commutation properties of disjunction are easily studied.
\begin{lemma} \label{le:comm_disj}
\mbox{$\forall \alpha, \beta, \gamma \in M$} that commute in pairs,
\mbox{$\alpha \vee \beta$} commutes with $\gamma$.
\end{lemma}
\begin{proof}
Obvious from Definition~\ref{def:disj} and Lemmas~\ref{le:comm_neg} 
and~\ref{le:comm_conj}.
\end{proof}

The following is easily proved: use Definition~\ref{def:disj}, {\bf Negation} 
and Lemmas~\ref{le:FPZ}, \ref{le:spec} and~\ref{le:commutation}.
\begin{lemma} \label{le:disj_Z}
For any commuting measurements, $\alpha$ and $\beta$,
their disjunction \mbox{$\alpha \vee \beta$} is the unique measurement
$\gamma$ such that \mbox{$Z(\gamma) = Z(\alpha) \cap Z(\beta)$}.
\end{lemma}

\begin{lemma} \label{le:disj}
If \mbox{$\alpha, \beta \in M$} {\em commute}, then
\mbox{$FP(\alpha) \cup FP(\beta) \subseteq FP(\alpha \vee \beta)$}.
\end{lemma}
The inclusion is, in general, strict.
\begin{proof}
Since \mbox{$Z(\alpha \vee \beta) \subseteq Z(\alpha)$}, 
by Lemma~\ref{le:FPZ}.
\end{proof}
Contrary to what holds in classical logic, in M-algebras we can have
a state $x$ that satisfies the disjunction $\alpha \vee \beta$ but
does not satisfy any one of $\alpha$ or $\beta$.
This is particularly interesting when $\alpha$ and $\beta$ represent
measurements of different values for the same physical quantity.
In this case, one is tempted to say that such an $x$ 
satisfies $\alpha$ not entirely but in {\em part} and $\beta$ in some other
part. In the Hilbertian formalism $x$ is a linear combination of the two 
vectors $\alpha(x)$ and $\beta(x)$: 
\mbox{$x = c_{1} \alpha(x) + c_{2} \beta(x)$}. The coefficients $c_{1}$
and $c_{2}$ describe in what proportions the state $x$, that satisfies
$\alpha \vee \beta$ satisfies $\alpha$ and $\beta$ respectively.
The consideration of structures richer than M-algebras that include this
quantitative information is left for future work.

\subsubsection{Implication} \label{sec:implication}
Implication ($\rightarrow$) is probably the most interesting connective.
It will play a central role in our treatment of connectives.
\begin{definition} \label{def:impl}
For any {\em commuting}
measurements \mbox{$\alpha, \beta \in M$}, the implication 
\mbox{$\alpha \rightarrow \beta$} is defined by:
\mbox{$\alpha \rightarrow \beta =$} 
\mbox{$\neg ( \alpha \wedge \neg \beta)$}. 
\end{definition}
By Corollary~\ref{le:co_neg}, the measurements
$\alpha$ and $\neg \beta$ commute, therefore their conjunction is
well-defined and the definition of implication is well-formed.

The commutation properties of implication are easily studied.
\begin{lemma} \label{le:comm_impl}
\mbox{$\forall \alpha, \beta, \gamma \in M$} that commute in pairs,
\mbox{$\alpha \rightarrow \beta$} commutes with $\gamma$.
\end{lemma}
\begin{proof}
Obvious from Definition~\ref{def:impl} and Lemmas~\ref{le:comm_neg} 
and~\ref{le:comm_conj}.
\end{proof}

The following is easily proved: use Definition~\ref{def:impl}, {\bf Negation} 
and Lemmas~\ref{le:FPZ}, \ref{le:spec} and~\ref{le:commutation}.
\begin{lemma} \label{le:impl_Z}
For any commuting measurements, $\alpha$ and $\beta$,
their implication \mbox{$\alpha \rightarrow \beta$} is the unique measurement
$\gamma$ such that \mbox{$Z(\gamma) = FP(\alpha) \cap Z(\beta)$}.
\end{lemma}

Lemma~\ref{le:impl_Z} characterizes the zeros of 
\mbox{$\alpha \rightarrow \beta$}.
Our next result characterizes the fixpoints of 
\mbox{$\alpha \rightarrow \beta$} in a most telling and useful way.
\begin{lemma} \label{le:impl_FP}
For any commuting measurements, $\alpha$ and $\beta$,
their implication \mbox{$\alpha \rightarrow \beta$} is the unique measurement
$\gamma$ such that 
\mbox{$FP(\gamma) =$} \mbox{$\{x \in X \mid \alpha(x) \in FP(\beta)\}$}.
\end{lemma}
\begin{proof}
Assume $\alpha$ and $\beta$ commute, and \mbox{$x \in X$}.
Now,
\mbox{$\alpha(x) \in$} \mbox{$FP(\beta)$} iff (by {\bf Negation})
\mbox{$\alpha(x) \in$} \mbox{$Z(\neg \beta)$} iff
\mbox{$(\alpha \circ (\neg \beta))(x) = 0$} iff (by Definition~\ref{def:conj}
and Lemma~\ref{le:comm_neg})
\mbox{$(\alpha \wedge (\neg \beta))(x) = 0$} iff
\mbox{$x \in$} \mbox{$Z(\alpha \wedge (\neg \beta))$} iff (by {\bf Negation})
\mbox{$x \in$} \mbox{$FP(\neg (\alpha \wedge (\neg \beta)))$} 
iff (by Definition~\ref{def:impl})
\mbox{$x \in$} \mbox{$FP(\alpha \rightarrow \beta)$}.
\end{proof}
The following is immediate.
\begin{corollary} \label{le:local_MP}
For any commuting measurements $\alpha$ and $\beta$,
if \mbox{$x \in FP(\alpha)$} and \mbox{$x \in FP(\alpha \rightarrow \beta)$},
then \mbox{$x \in FP(\beta)$}.
\end{corollary}

One may now ask whether the propositional connectives we have defined amongst
commuting measurements behave classically. In particular, assuming that
measurements $\alpha$, $\beta$ and $\gamma$ commute in pairs, 
does the distribution law hold, i.e., is it true that
\mbox{$(\alpha \vee \beta) \wedge \gamma =$}
\mbox{$(\alpha \wedge \gamma) \vee (\beta \wedge \gamma)$}.
In the next section, we shall show that amongst commuting measurements 
propositional connectives behave classically.

\section{Amongst commuting measurements connectives are classical}
\label{sec:classical_conn}
Let us, first, remark on the commutation properties described in 
Lemmas~\ref{le:comm_neg}, \ref{le:comm_conj}, \ref{le:comm_disj}
and~\ref{le:comm_impl}.
Those lemmas imply that, given any set \mbox{$A \subseteq M$} of measurements
in an M-algebra, such that any two elements of $A$ commute, one may consider
the propositional calculus built on $A$ (as atomic propositions).
Each such proposition describes a measurement in the original M-algebra
(an element of $M$) and all such measurements commute.
We shall denote by $Prop(A)$ the propositions built on $A$.

We shall now show that, in any such $Prop(A)$ 
all classical propositional tautologies hold at every state \mbox{$x \in X$}.
\begin{theorem} \label{the:class_calc}
Let \mbox{$\langle X , M \rangle$} be an M-algebra.
Let \mbox{$A \subseteq M$} be a set of pairwise {\em commuting} measurements.
If $\alpha \in Prop(A)$ is a classical propositional tautology, then
\mbox{$FP(\alpha) = X$}.
\end{theorem}
The converse does not hold, since it is easy to build M-algebras in which,
for example, a given measurement holds at every state.

We shall use the axiomatic system for propositional calculus found on p. 31
of Mendelson's~\cite{Mendelson:Intro} to prove that any classical tautology
$\alpha$ built out using only negation and implication 
has the property claimed.
We shall then show that conjunction and disjunction may be defined in terms
of negation and implication as usual.
The proof will proceed in six steps: Modus Ponens, the three axiom schemes
of Mendelson's system, conjunction and disjunction.
The reader should notice how tightly the three axiom schemas correspond to the
commutation assumption.

\begin{lemma} \label{le:MP}
For any commuting measurements $\alpha$ and $\beta$,
if \mbox{$FP(\alpha) = X$} and \mbox{$FP(\alpha \rightarrow \beta) = X$},
then \mbox{$FP(\beta) = X$}.
\end{lemma}
\begin{proof}
By Corollary~\ref{le:local_MP}.
\end{proof}

\begin{lemma} \label{le:Sch1}
For any commuting measurements $\alpha$ and $\beta$,
\[
FP(\alpha \rightarrow (\beta \rightarrow \alpha)) = X.
\]
\end{lemma}
\begin{proof}
Since $\alpha$ and $\beta$ commute, for any \mbox{$x \in X$}:
\mbox{$\beta(\alpha(x)) =$} \mbox{$\alpha(\beta(x))$}, 
therefore, by {\bf Idempotence}, we have
\mbox{$\beta(\alpha(x)) \in FP(\alpha)$}.
By Lemma~\ref{le:impl_FP}, for any $x$, 
\mbox{$\alpha(x) \in FP(\beta \rightarrow \alpha)$}.
By the same lemma:
\mbox{$x \in FP(\alpha \rightarrow (\beta \rightarrow \alpha))$}.
\end{proof}

\begin{lemma} \label{le:Sch2}
For any pairwise commuting measurements $\alpha$, $\beta$ and $\gamma$
\[
FP((\alpha \rightarrow (\beta \rightarrow \gamma)) \rightarrow
((\alpha \rightarrow \beta) \rightarrow (\alpha \rightarrow \gamma))) = X.
\]
\end{lemma}
\begin{proof}
By Lemma~\ref{le:impl_FP}, it is enough to show
that for any \mbox{$x \in X$}, if
\mbox{$y =$} \mbox{$(\alpha \rightarrow (\beta \rightarrow \gamma))(x)$}, then,
if we define \mbox{$z =$} \mbox{$(\alpha \rightarrow \beta)(y)$}, and define
\mbox{$w =$} \mbox{$\alpha(z)$}, then we have:
\mbox{$\gamma(w) =$} \mbox{$w$}.
But since all the measurements above commute, by {\bf Idempotence},
the state $w$ satisfies \mbox{$\alpha \rightarrow (\beta \rightarrow \gamma)$},
\mbox{$\alpha \rightarrow \beta$} and $\alpha$.
By Corollary~\ref{le:local_MP}, $w$ satisfies $\beta$ and 
\mbox{$\beta \rightarrow \gamma$}. 
For the same reason $w$ satisfies $\gamma$.
\end{proof}

\begin{lemma} \label{le:Sch3}
For any commuting measurements $\alpha$ and $\beta$,
\[
FP((\neg \beta \rightarrow \neg \alpha) \rightarrow 
((\neg \beta \rightarrow \alpha) \rightarrow \beta)) = X.
\]
\end{lemma}
\begin{proof}
By Lemma~\ref{le:impl_FP}, it is enough to show
that for any \mbox{$x \in X$}, if
\mbox{$y =$} \mbox{$(\neg \beta \rightarrow \neg \alpha)(x)$}, then,
if we define \mbox{$z =$} \mbox{$(\neg \beta \rightarrow \alpha)(y)$} 
then we have:
\mbox{$\beta(z) =$} \mbox{$z$}.
But since all the measurements above commute, by {\bf Idempotence},
the state $z$ satisfies \mbox{$\neg \beta \rightarrow \neg \alpha$} and
\mbox{$\neg \beta \rightarrow \alpha$}.
Therefore, by Lemma~\ref{le:impl_FP}, $(\neg \beta)(z)$ satisfies
both $\neg \alpha$ and $\alpha$. Therefore \mbox{$(\neg \beta)(z) = 0$}
and therefore, by {\bf Negation}, \mbox{$z \in FP(\beta)$}.
\end{proof}

\begin{lemma} \label{le:conj_class}
For any commuting measurements $\alpha$ and $\beta$,
\mbox{$\alpha \wedge \beta = \neg (\alpha \rightarrow \neg \beta)$}.
\end{lemma}
\begin{proof}
\[
FP(\neg (\alpha \rightarrow \neg \beta)) = Z(\alpha \rightarrow \neg \beta) =
FP(\alpha) \cap Z(\neg \beta) = FP(\alpha) \cap FP(\beta).
\]
By {\bf Negation}, Lemma~\ref{le:impl_Z} and {\bf Negation}.
The conclusion then follows from Lemma~\ref{le:unique_conj}.
\end{proof}

\begin{lemma} \label{le:disj_class}
For any commuting measurements $\alpha$ and $\beta$,
\mbox{$\alpha \vee \beta = (\neg \alpha) \rightarrow \beta$}.
\end{lemma}
\begin{proof}
\[
Z((\neg \alpha) \rightarrow \beta) = FP(\neg \alpha) \cap Z(\beta) =
Z(\alpha) \cap Z(\beta).
\]
By Lemma~\ref{le:impl_Z} and {\bf Negation}.
The conclusion then follows from Lemma~\ref{le:disj_Z}.
\end{proof}

We have proved Theorem~\ref{the:class_calc}.

The following is a Corollary.
\begin{corollary} \label{co:implication}
Let \mbox{$\langle X , M \rangle$} be an M-algebra.
Let \mbox{$A \subseteq M$} be a set of pairwise {\em commuting} measurements.
If $\alpha, \beta \in Prop(A)$ are such that $\alpha$ logically implies 
$\beta$, i.e., \mbox{$\alpha \models \beta$}, then
\mbox{$FP(\alpha) \subseteq FP(\beta)$}.
If $\alpha$ and $\beta$ are logically equivalent, they are equal.
\end{corollary}
\begin{proof}
Since \mbox{$\alpha \rightarrow \beta$} is a tautology, by 
Theorem~\ref{the:class_calc}, \mbox{$FP(\alpha \rightarrow \beta) = X$}.
By Lemma~\ref{le:impl_FP}, then, \mbox{$FP(\alpha) \subseteq FP(\beta)$}.
If $\alpha$ and $\beta$ are logically equivalent, they have the same set
of fixpoints by the above, and, by Lemma~\ref{le:spec} they are equal.
\end{proof}
\section{Orthomodularity of M-algebras} \label{sec:orthomodularity}
In this section we shall clear up the relation between our M-algebras and 
the lattice structures largely studied previously (see~\cite{Redei:Algebraic}
for an in-depth study and review).
The set of measurements $M$ of an M-algebra is naturally endowed with a partial
order.
\begin{definition} \label{def:order}
Let \mbox{$\langle X , M \rangle$} be an M-algebra. For any 
\mbox{$\alpha, \beta \in M$}, let \mbox{$\alpha \leq \beta$} iff 
\mbox{$FP(\alpha) \subseteq FP(\beta)$} (or equivalently, 
by Lemma~\ref{le:FPZ}, iff \mbox{$Z(\beta) \subseteq Z(\alpha)$}).
\end{definition}
\begin{lemma} \label{le:lattice}
Let \mbox{$\langle X , M \rangle$} be an M-algebra.
The relation $\leq$ on $M$ is a partial order. 
The measurement $\top$ is the top element and the measurement $\bot$ 
is the bottom element, i.e.:
for any \mbox{$\alpha \in M$}, \mbox{$\bot \leq \alpha \leq \top$}.
Any two {\em commuting}
measurements have a greatest lower bound and a least upper bound in $M$.
\end{lemma}
The set $M$ is not, in general, a lattice, under $\leq$.
\begin{proof}
The relation $\leq$ is a partial order because $\subseteq$ is.
For any $\alpha \in M$, \mbox{$FP(\bot) =$}
\mbox{$\{0\} \subseteq$} \mbox{$FP{\alpha} \subseteq$}
\mbox{$X = FP(\top)$}.
Consider any two commuting measurements $\alpha$ and $\beta$.
By Lemma~\ref{le:unique_conj} the measurement \mbox{$\alpha \wedge \beta$} is
the greatest lower bound of $\alpha$ and $\beta$.
By Lemma~\ref{le:disj_Z} and the definition of $\leq$ in terms of $Z$ it is
clear that \mbox{$\alpha \vee \beta$} is the least upper bound of
$\alpha$ and $\beta$.
\end{proof}

M-algebras represent a departure from the structures previously considered by
researchers in Quantum Logic because they are not lattices.
Orthomodular lattices have been considered by most to be the structure of 
choice. Orthomodular lattices are lattices equipped with a unary operation
of orthocomplementation ${ }^{\bot}$ satisfying the following properties
(see for example~\cite{Redei:Algebraic}, pages 33-35).
For any \mbox{$\alpha, \beta, \gamma$}:
\begin{enumerate}
\item \label{doubleneg} \mbox{$(\alpha^{\bot})^{\bot} = \alpha$},
\item \label{inversion} if \mbox{$\alpha \leq \beta$}, then 
\mbox{$\beta^{\bot} \leq \alpha^{\bot}$},
\item \label{aandnota} 
the greatest lower bound of $\alpha$ and $\alpha^{\bot}$ is the bottom
element, 
\item \label{aornota} 
the least upper bound of $\alpha$ and $\alpha^{\bot}$ is the top element, and
\item \label{short_orthomodularity} 
if \mbox{$\alpha \leq \beta$} and
$\gamma$ is the greatest lower bound of $\alpha^{\bot}$ and $\beta$, then
$\beta$ is the least upper bound of $\gamma$ and $\alpha$.
\end{enumerate}
\begin{lemma} \label{le:orthononlattice}
Let \mbox{$\langle X , M \rangle$} be an M-algebra.
Each one of the properties above holds for the partially ordered set 
\mbox{$\langle M , \leq \rangle$} when negation is taken for 
orthocomplementation.
\end{lemma}
\begin{proof}
Item~\ref{doubleneg} holds by Corollary~\ref{le:double_neg}.
Item~\ref{inversion} holds by {\bf Negation} and Lemma~\ref{le:FPZ}.
Item~\ref{aandnota} holds since $\alpha$ and $\neg \alpha$ commute,
by Lemma~\ref{le:lattice}, their greatest lower bound is
\mbox{$\alpha \wedge \neg \alpha$} and 
\mbox{$FP(\alpha \wedge \neg \alpha) =$}
\mbox{$FP(\alpha) \cap Z(\alpha) = \{0\ =$}
\mbox{$FP(\bot)$}.
Item~\ref{aornota} holds similarly:
\mbox{$Z(\alpha \vee \not \alpha) = \{0\} = Z(\top)$}.
Item~\ref{short_orthomodularity} follows from the following considerations.
If \mbox{$\alpha \leq \beta$}, then $\alpha$ and $\beta$ commute by 
Lemma~\ref{le:incl}.
Therefore, by Lemma~\ref{le:comm_neg}, $\neg \alpha$ and $\beta$ commute,
and by Lemma~\ref{le:lattice} \mbox{$\gamma = \neg \alpha \wedge \beta$}. 
Also, $\alpha$ and $\gamma$ commute by Lemma~\ref{le:comm_disj} and
by Lemma~\ref{le:lattice} all we have to show is that
\mbox{$beta = \alpha \vee (\neg \alpha \wedge \beta$}.
But $\beta$ is logically equivalent to 
\mbox{$(\alpha \wedge \beta) \vee (\neg \alpha \wedge \beta)$}.
Corollary~\ref{co:implication} implies that 
\mbox{$b = (\alpha \wedge \beta) \vee (\neg \alpha \wedge \beta)$}.
By assumption \mbox{$\alpha \leq \beta$} and, by Lemma~\ref{le:incl}
\mbox{$\alpha \wedge \beta = \alpha$}.
We conclude 
that \mbox{$b =$} \mbox{$\alpha \vee (\neg \alpha \wedge \beta)$}.  
\end{proof}

The next Section will consider separable and strongly-separable M-algebras.

\section{Separable and stronly-separable M-algebras} 
\label{sec:separable_andstrong}
\subsection{Separable M-algebras} \label{sec:separable}
\begin{lemma} \label{le:class_commute}
In a {\em separable} M-algebra, a measurement is classical 
if and only if it commutes with any measurement.
\end{lemma}
\begin{proof}
Suppose $\alpha$ is classical.
Consider any \mbox{$x \in X$} and any \mbox{$\beta \in M$}.
Since $\alpha$ is classical we know that 
\mbox{$x \in$} \mbox{$FP(\alpha)$} or \mbox{$x \in$} \mbox{$Z(\alpha)$}
and \mbox{$\beta(x) \in$} \mbox{$FP(\alpha)$} or
\mbox{$\beta(x) \in$} \mbox{$Z(\alpha)$}.
If \mbox{$x \in$} \mbox{$FP(\alpha)$}, by Lemma~\ref{le:definiteness}, 
\mbox{$\beta(x) \in$} \mbox{$Z(\alpha)$} implies \mbox{$x \in$}
\mbox{$Z(\beta)$}and 
\mbox{$(\alpha \circ \beta)(x) =$} \mbox{$0 =$}
\mbox{$(\beta \circ \alpha)(x)$}.
But \mbox{$\beta(x) \in$} \mbox{$FP(\alpha)$} implies
\mbox{$(\alpha \circ \beta)(x) =$}
\mbox{$\beta(x) =$}
\mbox{$(\beta \circ \alpha)(x)$}.

If \mbox{$x \in$} \mbox{$Z(\alpha)$} and 
\mbox{$\beta(x) \in$} \mbox{$FP(\alpha)$}, 
by Lemma~\ref{le:Def_dual}, \mbox{$\beta(x) = 0$} and
\mbox{$(\alpha \circ \beta)(x) =$} \mbox{$0 =$}
\mbox{$(\beta \circ \alpha)(x)$}.
If \mbox{$\beta(x) \in$}  \mbox{$Z(\alpha)$}, then
\mbox{$(\alpha \circ \beta)(x) =$}
\mbox{$0 =$}
\mbox{$(\beta \circ \alpha)(x)$}.

Suppose, now that $\alpha$ commutes with any measurement $\beta$.
By contradiction, assume \mbox{$\alpha(x) \neq 0$} and 
\mbox{$\alpha(x) \neq x$}. By {\bf Separability} there is some
measurement $\gamma$ such that \mbox{$x \in FP(\gamma)$}
and \mbox{$\alpha(x) \not \in FP(\gamma)$}.
But $\alpha$ and $\gamma$ commute and:
\mbox{$(\alpha \circ \gamma)(x) =$}
\mbox{$(\gamma \circ \alpha)(x) =$}
\mbox{$\alpha(x)$}.
We see that \mbox{$\alpha(x) \in FP(\gamma)$}, a contradiction.
\end{proof}

Note that a measurement $\alpha$ is classical 
(see Definition~\ref{def:classical}) iff \mbox{$Def(\alpha) = X$}.
\begin{lemma} \label{le:connectives_class}
If $\alpha$ is classical, so is $\neg \alpha$.
If $\alpha$ and $\beta$ are classical, then so are 
\mbox{$\alpha \wedge \beta$}, \mbox{$\alpha \vee \beta$} 
and \mbox{$\alpha \rightarrow \beta$}.
\end{lemma}
\begin{proof}
If $\alpha$ is classical, \mbox{$Def(\alpha) = X$} and therefore
\mbox{$Def(\neg \alpha) = X$}.
For conjunction \mbox{$(\alpha \wedge \beta)(x) =$}
\mbox{$(\alpha \circ \beta)(x) =$}
\mbox{$(\beta \circ \alpha)(x)$}.
If either $\alpha(x)$ or $\beta(x)$ is $0$ then 
\mbox{$(\alpha \wedge \beta)(x) = 0$}, otherwise
\mbox{$\alpha(x) = x = \beta(x)$} and 
\mbox{$(\alpha \wedge \beta)(x) = x$}.
The definitions of disjunction and implication in terms of negation and
conjunction, then ensure the claim.
\end{proof}

\subsection{Strongly-separable M-algebras} \label{sec:strongly_sep}
We shall show that much of the linear dependency structure of orthomodular
and Hilbert spaces is present in any strongly-separable M-algebra.
Theorem~\ref{the:proj} shows that the action of the measurements in
a strongly-separable M-algebra is already encoded in its order structure.
\begin{theorem} \label{the:proj}
Let \mbox{$\langle X, M \rangle$} be any strongly-separable M-algebra.
The following two properties are satisfied for any \mbox{$\alpha \in M$} and
any \mbox{$x \in X$}:
\begin{enumerate} 
\item \label{dependence}
\mbox{$x \in FP(\alpha \rightarrow e_{\alpha(x)})$}, and
\item \label{uniqueness}
if \mbox{$x \not \in Z(\alpha)$}, 
there exists a unique \mbox{$y \in FP(\alpha)$} such that
\mbox{$x \in FP(\alpha \rightarrow e_{y})$}. This $y$ is $e_{\alpha(x)}$.
\end{enumerate}
\end{theorem}
\begin{proof}
Note, first that the measurements \mbox{$\alpha \rightarrow e_{\alpha(x)}$}
and \mbox{$\alpha \rightarrow e_{y}$} are well-defined since $\alpha$ and
$e_{\alpha(x)}$ commute by Lemma~\ref{le:incl} since
\mbox{$FP(e_{\alpha(x)}) = \{0 , x\} \subseteq FP(\alpha)$} by Idempotence, 
and, similarly $\alpha$ and $e_{y}$ commute if \mbox{$y \in FP(\alpha)$}.

Both claims follow straightforwardly from Lemma~\ref{le:impl_FP} and
the following: \mbox{$\alpha(x) \in FP(e_{\alpha(x)})$}, and
\mbox{$\alpha(x) \in FP(e_{y})$} iff \mbox{$\alpha(x) = 0$} or
\mbox{$\alpha(x) = y$}.
\end{proof}
The following shows the decomposition of any state in its orthogonal 
components.
\begin{corollary} \label{co:linear}
Let \mbox{$\langle X, M \rangle$} be any strongly-separable M-algebra.
For any \mbox{$\alpha \in M$} and
any \mbox{$x \in X$}, 
\mbox{$x \in FP(e_{\alpha(x)} \vee e_{(\neg \alpha)(x)})$}.
\end{corollary}
\begin{proof}
Note that, since $e_{\alpha(x)}$ and $e_{(\neg \alpha)(x)}$ are orthogonal, 
they commute and the disjunction of the claim is well defined.

By Theorem~\ref{the:proj}, we have both
\mbox{$x \in FP(\neg \alpha \vee e_{\alpha(x)})$} and
\mbox{$x \in FP(\alpha \vee e_{(\neg \alpha)(x)})$}.
But all measurements mentioned above commute and the conjunction of the
two disjunctions is well-defined and
\mbox{$x \in FP((\alpha \vee e_{(\neg \alpha)(x)}) \wedge 
(\neg \alpha \vee e_{\alpha(x)}))$}.
But, by Theorem~\ref{the:class_calc}:
\[
(\alpha \vee e_{(\neg \alpha)(x)}) \wedge 
(\neg \alpha \vee e_{\alpha(x)}) =
e_{(\neg \alpha)(x)} \vee e_{\alpha(x)}
\]
\end{proof}

\section{Reflections and further work}
The formalism of M-algebras is weaker 
than that of Hilbert spaces and may be motivated by epistemological
concerns. Nevertheless some of the properties of 
quantum measurements may be understood in this weaker formalism.
Should additional principles be incorporated into M-algebras?
Probably, the next step will be to incorporate some quantitative information
about the relation between a state and a measurement.

Corollary~\ref{le:class_commute} may explain why elementary particles
all have a definite value for their total spin. 
The corresponding Hermitian operator commutes with every spin operator,
and most probably any physically meaningful operator on an isolated
particle. Therefore the corresponding measurements (of the different
values of the total spin) are classical. This means that, for any elementary
particle its total spin has a definite value. Particles with different
definite values for their total spin are better considered different
particles: no measurement can change their total spin.

Can the formalism of M-algebras throw light on superselection rules,
such as the symmetrization postulate? 

Another tantalizing question concerns the tensor product of M-algebras.
What should it be? Would this explain why a quantic system composed of two
separate subsystems must be represented by the tensor product of the spaces
representing the subsystems?

\bibliographystyle{plain}

\end{document}